\documentclass[twocolumn, switch]{article} 

\usepackage{arxiv}

\usepackage{amsmath, amsthm, amssymb, amsfonts}
\usepackage{algpseudocode}
\usepackage{algorithm}

\usepackage[numbers,square]{natbib}

\usepackage[utf8]{inputenc}	
\usepackage[T1]{fontenc}	
\usepackage{xcolor}		
\usepackage[colorlinks = true,
            linkcolor = purple,
            urlcolor  = blue,
            citecolor = cyan,
            anchorcolor = black]{hyperref}	
\usepackage{booktabs} 		
\usepackage{nicefrac}		
\usepackage{microtype}		
\usepackage{lineno}		
\usepackage{float}			

\usepackage{lipsum}		
\usepackage{subcaption}

\usepackage{newfloat}
\DeclareFloatingEnvironment[name={Supplementary Figure}]{suppfigure}
\usepackage{sidecap}
\sidecaptionvpos{figure}{c}

\usepackage{titlesec}
\titlespacing\section{0pt}{12pt plus 3pt minus 3pt}{1pt plus 1pt minus 1pt}
\titlespacing\subsection{0pt}{10pt plus 3pt minus 3pt}{1pt plus 1pt minus 1pt}
\titlespacing\subsubsection{0pt}{8pt plus 3pt minus 3pt}{1pt plus 1pt minus 1pt}

\usepackage{tikz,xcolor,hyperref}

\definecolor{lime}{HTML}{A6CE39}
\DeclareRobustCommand{\orcidicon}{
	\begin{tikzpicture}
	\draw[lime, fill=lime] (0,0) 
	circle [radius=0.16] 
	node[white] {{\fontfamily{qag}\selectfont \tiny ID}};
	\draw[white, fill=white] (-0.0625,0.095) 
	circle [radius=0.007];
	\end{tikzpicture}
	\hspace{-2mm}
}
\foreach \x in {A, ..., Z}{\expandafter\xdef\csname orcid\x\endcsname{\noexpand\href{https://orcid.org/\csname orcidauthor\x\endcsname}
			{\noexpand\orcidicon}}
}

\title{Towards More Accessible Scientific PDFs for People with Visual Impairments: Step-by-Step PDF Remediation to Improve Tag Accuracy}

\usepackage{authblk}

\author[1, 2\thanks{\tt{scmx@zhaw.ch}}]{Felix M. Schmitt-Koopmann\orcidA{}}
\author[2]{Elaine M. Huang\orcidB{}}
\author[1]{Hans-Peter Hutter\orcidC{}}
\author[1]{Alireza Darvishy\orcidE{}}

\affil[1]{Institute of Computer Science, ZHAW, 8401 Winterthur, Switzerland}
\affil[2]{People and Computing Lab, University of Zurich, 8050 Zurich, Switzerland}

\begin{document}

\twocolumn[ 
  \begin{@twocolumnfalse} 
  
\maketitle

\begin{abstract}
PDF inaccessibility is an ongoing challenge that hinders individuals with visual impairments from reading and navigating PDFs using screen readers. This paper presents a step-by-step process for both novice and experienced users to create accessible PDF documents, including an approach for creating alternative text for mathematical formulas without expert knowledge. In a study involving nineteen participants, we evaluated our prototype PAVE 2.0 by comparing it against Adobe Acrobat Pro, the existing standard for remediating PDFs. Our study shows that experienced users improved their tagging scores from 42.0\% to 80.1\%, and novice users from 39.2\% to 75.2\% with PAVE 2.0. Overall, fifteen participants stated that they would prefer to use PAVE 2.0 in the future, and all participants would recommend it for novice users. Our work demonstrates PAVE 2.0's potential for increasing PDF accessibility for people with visual impairments and highlights remaining challenges.
\end{abstract}
\vspace{0.35cm}

  \end{@twocolumnfalse} 
] 

\section{Introduction}
\label{sec:introduction}
Many people with visual impairments rely on screen readers to access and read PDFs. Unfortunately, research has shown that only a tiny percentage of the trillions of PDF documents available are accessible for individuals who use screen readers and meet the PDF Universal Access (UA) standard \cite{wang_improving_2021}. This means that people with visual impairments cannot properly read most PDFs with screen readers, posing a significant problem — especially in STEM fields, where PDF is the dominant format \cite{bigham_uninteresting_2016}. Consequently, 2.4\% of the US population is not able to easily read most scientific PDFs \cite{national_federation_of_the_blind_blindness_2017}. This presents an extra barrier for people with visual impairment wishing to pursue studies or careers in STEM fields \cite{fichten_higher_2020} which contributes to the underrepresentation of people with disabilities in research \cite{sills_disability_2019}. 

The importance of this issue is underscored by several laws and resolutions created to ensure access to public documents for everybody. The oldest of these is the US Rehabilitation Act Section 508 of 1998 \cite{us_congress_section_1998}. It requires US federal departments and agencies to make electronic and information technology accessible to people with disabilities. Moreover, the 2008 United Nations Convention on the Rights of Persons with Disabilities \cite{un_general_assembly_convention_2007} and the European Accessibility Act \cite{eu_directive_2019} of 2019 require that critical products and services be usable by people with disabilities.

The discrepancy between these legal regulations and the continued lack of accessibility of PDFs documents can be attributed in large part to the lack of awareness regarding PDF accessibility and the challenge of making PDF documents accessible. The CHI conferences have made significant efforts to promote accessibility in CHI papers by providing templates, guidelines, and services \cite{bigham_uninteresting_2016}. However, the current processes for making PDFs accessible are still time-consuming and error-prone, and require expert knowledge \cite{jembu_rajkumar_pdf_2020}, challenges that many authors have likely experienced. As a result, our analysis reveals that conference papers often include tags, but the accuracy of the tags need to be improved.

Hence, there is a need for methods and tools that allow people to remediate PDFs with a high degree of accessibility \cite{bigham_uninteresting_2016, jembu_rajkumar_pdf_2020, pradhan_development_2022}. Such tools should not require that individuals have substantial specialized knowledge or invest large amounts of time to use them successfully. 
We developed a novel PDF remediation process with eight distinct steps to prevent users from becoming overwhelmed by the complexities of PDF accessibility details. One of our goals was to understand how our method affects the PDF remediation process from the user perspective, as well as how it affects the accessibility of the PDFs themselves. To evaluate our process, we conducted a user study involving nineteen participants with varying levels of experience in PDF remediation. 
Participants remediated a PDF once with our prototype and once with the industry standard tool, Adobe Acrobat Pro. It showed that our process allows novice as well as experienced users to increase the tag accuracy of their PDFs by around 90\% and fifteen of nineteen participants stated that they would like to use our tool in the future. 

In this work, we present the following major contributions: 1) A user study to evaluate our step-by-step PDF remediation process, 2) a novel AI-based method to automatically create alternative texts for mathematical formulas, and 3) a novel accessibility score with thirteen criteria to reliably evaluate the tag accuracy of a PDF. 

The remainder of the paper is organized as follows: Section \ref{sec:RelatedWork} explores the related work. Section \ref{sec:InterfaceDesign} presents the developed process and our prototype, PAVE 2.0. Section \ref{sec:UserStudyDesign} presents the design of the user test. The results of the user test are presented in Section \ref{sec:Results} and we discuss the results in Section \ref{sec:Discussion}. Lastly, Section \ref{sec:Conclusion} contains concluding remarks.

\section{Related Work}
\label{sec:RelatedWork}
An accessible PDF can be created in two ways. The first method involves exporting an accessible PDF directly from the document editing software, provided that the software has an accessible PDF export option. However, not all document editors have this feature, and the author has to follow specific guidelines when creating the document to ensure that it is exported correctly as an accessible PDF. The second option is PDF remediation in which the author modifies an existing PDF to make it accessible. We focus on the PDF remediation method in this work because it is applicable regardless of the process used to create the PDF originally. This process is therefore valuable for making new PDFs accessible at time of their creation, but also for potentially making the vast quantity of inaccessible PDFs already in existence accessible to individuals who use screen readers as well. However, the complexity of the PDF remediation process has contributed directly to the current situation in which hardly any scientific PDFs are accessible, as the following research shows.

Wang et \textit{al}. \cite{wang_improving_2021} automatically assessed over 11,000 scientific PDFs, published between 2010 and 2019, and found that only 2.4\% of the PDFs satisfied all their accessibility criteria. Nganji et \textit{al}. \cite{nganji_assessment_2018} manually analyzed 200 articles from four journals between 2014 and 2018. They found that only 15.5\% of the documents contained tags. Darvishy et \textit{al}. \cite{darvishy_state_2023} investigated 2,500 papers in repositories of five German-speaking universities in Switzerland from 2018 to 2022 in a semi-automatic way and found that only 11.5\% of the papers contained tags. Pierrès et \textit{al}. \cite{pierres_pdf_2024} examined 8,000 papers from four large scientific repositories mainly from 2023 and 2024 and found that papers that contained at least one tag and were marked as tagged comprised only 0.9\% of the papers for Wiley and 16.2\% for Elsevier. Even though the accessibility of scientific literature has improved in recent years, the vast majority of research literature is still not accessible to everyone.

This lack of document accessibility is not only a problem in regard to scientific literature. Drye et \textit{al}. \cite{drye_professionals_2023} conducted a study to understand the accessibility of business communication materials. The study revealed that one-third of the participants could not define what accessibility of documents means. Additionally, half of the participants did not know how to create an accessible document. Similar to the findings of Rajkumar et \textit{al}. \cite{jembu_rajkumar_pdf_2020}, this highlights how the lack of awareness of accessibility is one of the key reasons why documents are not made accessible.

Bigham et \textit{al}. \cite{bigham_uninteresting_2016} presented their experience when trying to make conference PDFs accessible. They highlighted that current tools for this task are too complicated and not usable for inexperienced users.
Similarly, Rajkumar et \textit{al}. \cite{jembu_rajkumar_pdf_2020} found that most researchers and practitioners in STEM fields are unhappy with the existing tools. For this reason, user-friendly tools for creating accessible documents are a major necessity to improve the accessibility of scientific and other literature.

The related work on methods for making PDFs accessible can be categorized into three groups. The first group involves converting PDFs into a more accessible format. For instance, Wang et \textit{al}. \cite{wang_improving_2021} developed SciA11y, a method to transform a PDF into an accessible HTML. This is especially beneficial for making existing PDFs accessible without manual work. However, conversion errors do occur and are not detected which is problematic. Besides SciA11y, there are many other tools available that can convert PDFs into HTML, as shown in the evaluation by Pathirana et \textit{al}. \cite{pathirana_comparative_2023}.

The second group considers how to check PDF accessibility. One of the most popular tools for this is the PDF Accessibility Checker (PAC) \cite{Uebelbacher_PDF_2014}. PAC allows users to analyze and reveal accessibility issues in a PDF but these cannot be corrected directly in the tool. Even more important, checkers are limited to machine-testable criteria which do not cover all important accessibility aspects, e.g., the correct tagging of headers or the reading order \cite{kumar_uncovering_2024}.

The third group comprises interactive tools that allow users to make PDFs accessible (PDF remediation). The most popular tool is Adobe Acrobat Pro \cite{adobe_adobe_2024}. Other professional PDF remediation tools are, e.g., CommonLook PDF \cite{commonlook_commonlook_2024}, Axes4 PDF \cite{axes4_axespdf_2024}, and Foxit Reader \cite{foxit_foxit_2024}. However, the workflows between these tools differ only slightly. They utilize an accessibility checker as a guide for the user, and the user can edit the structure tree directly. Additionally, they offer an automatic tagging feature that creates tags for the entire document.

Doblies et \textit{al}. \cite{doblies_pave_2014} developed the web application PAVE in 2014 upon which our prototype (PAVE 2.0) is based. The idea of PAVE was to provide a simple, semi-automatic process for non-experts to remediate PDFs. A comparison study with Adobe Acrobat Pro \cite{jembu_rajkumar_pdf_2020} revealed that a semi-automatic tool like PAVE is necessary to foster PDF accessibility, but PAVE at that time did not meet user expectations regarding intuitiveness or user experience. They found that a big part of the user frustration was that the participants did not know what was auto-tagged by PAVE and how to override it. Pradhan et \textit{al}. \cite{pradhan_development_2022} developed the Ally prototype to improve the user experience of remediation tools using best practices from HCI research. The major improvement to existing tools was that the user no longer interacted directly with the PDF structure tree which represents the logical structure of the content of a PDF. Instead, Ally splits the process into multiple logical subtasks to give the user more guidance. Their prototype presented user interfaces for four subtasks (regions, headers, reading order, and tables), but not for the complete PDF remediation process. This work incorporates the findings of Pradhan et \textit{al}. and extends it to a complete AI-supported PDF remediation process. 

\section{Interface Design of PAVE 2.0}
\label{sec:InterfaceDesign}
Our goal was to develop a semi-automatic process that makes the PDF remediation process more intuitive and easier while avoiding direct user interaction with the complex PDF structure tree. Inspired by Pradhan et \textit{al}. \cite{pradhan_development_2022}, we developed a PDF remediation process with eight steps.

Four of these steps (Region, Reading Order, Heading Structure, and Meta Information) are necessary for every PDF, regardless of its content. We decided to split the content-independent tasks into four distinct steps to ensure that each step focuses on a single, clearly defined task. The remaining four steps (Tables, Lists, Figures, and Mathematical Formulas) are content-dependent, reflecting their common presence in scientific PDFs. Although additional content-dependent steps could be considered, we identified these four steps as the most relevant for scientific papers. The final PDF tag structure is then constructed by combining the information from each of these steps.

This step-by-step approach ensures a logical workflow and prevents the user from creating invalid tags or invalid nesting of tags in the PDF structure tree. In contrast to most PDF remediation tools, we decided not to integrate an accessibility report or check in the prototype as it may foster a misleading impression of accessibility \cite{kumar_uncovering_2024}. We want to note that the first three steps (Region, Reading Order, Heading Structure) are inspired by Pradhan et \textit{al}. \cite{pradhan_development_2022}.

The user interface for each of the steps is split into three parts (see Figure \ref{fig:step1}). On the left side is the navigation pane for all eight steps. The workspace for each step is located in the center pane. On the right side, the page view shows the current PDF page with the tagged regions. At upper left of the workspace area there is a brief description of what the user should do in this step, what the page view is showing, and which features are available.

The page view always shows one rendered page of the PDF at a time with status information at the top, depending on the step. The user can zoom in and out of the page. Depending on the step, the user can hide different aspects of the visualization to reduce visual clutter and cognitive load \cite{tsurukawa_filtering_2015}.

The user interface was developed using the ReactJS library \cite{meta_platforms_react_2023}. It was designed with accessibility as a priority, e.g., using a high-contrast color palette. However, we overlooked a problematic red-green color combination in Step 2 which a participant highlighted later. While the prototype is not fully accessible (particularly the drawing functionalities) to streamline development, achieving complete accessibility remains essential for the production version. The PDF is modified and parsed on the back end, which is based on the back end of the original PAVE tool \cite{zhaw_init_pave_2023}.
In the following section, we explain each of the eight steps of the PAVE 2.0 remediation process.

\subsection{Step 1 - Regions}
\begin{figure*}
    \centering
    \includegraphics[width=\textwidth]{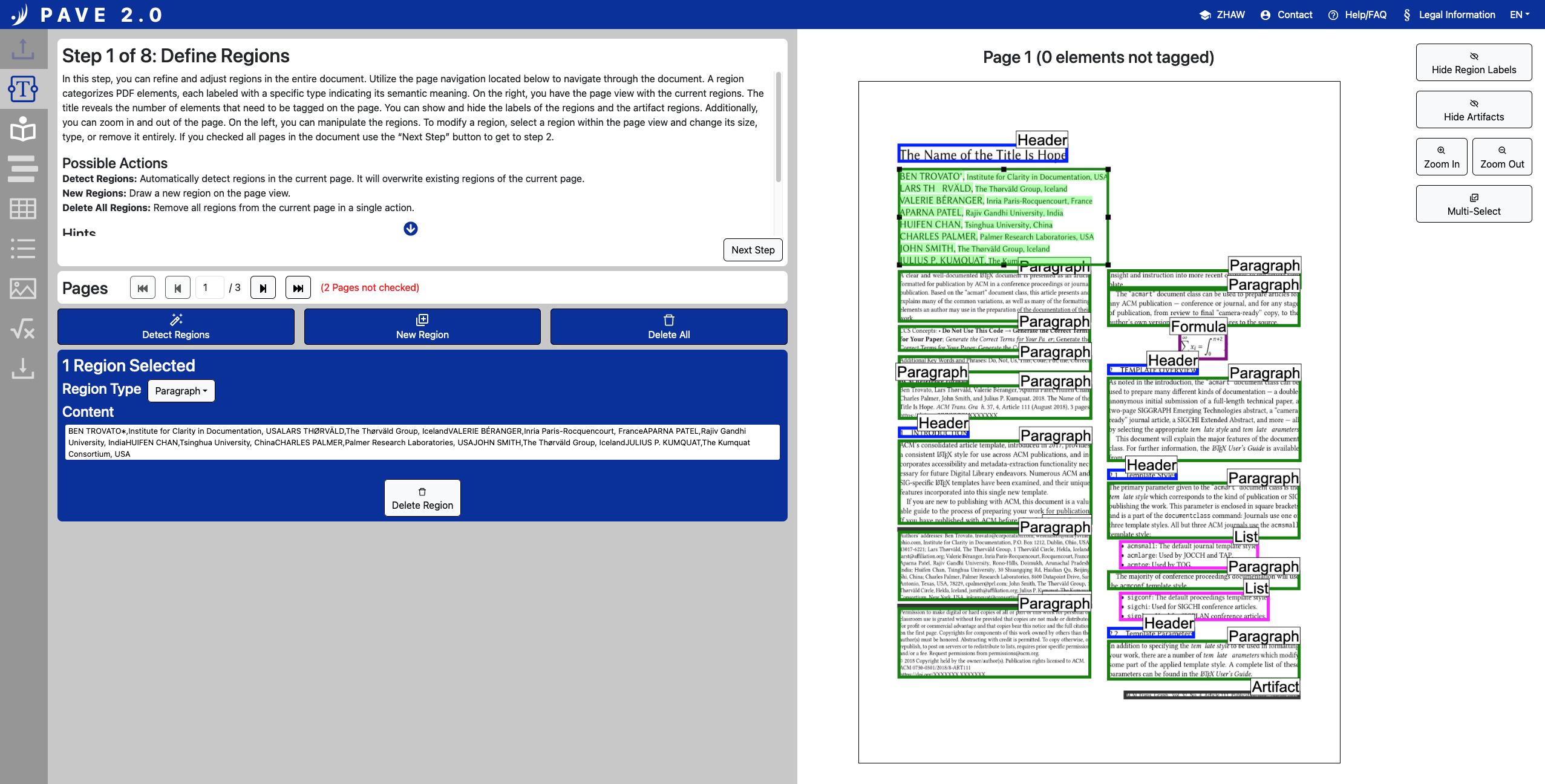}
    \caption{Screenshot of Step 1 (regions) in PAVE 2.0. The interface contains three interface panes from left to right: step navigation, workspace, page view.}
    \label{fig:step1}
\end{figure*}

In the first step of the PAVE 2.0 process, the user is required to identify the different regions on each page of the PDF, such as paragraphs, headings, lists, formulas, figures, captions, artifacts, and tables (see Figure \ref{fig:step1}). Users are provided with two options to define regions. Firstly, users can select untagged content such as text, images, or drawing elements (PDF operators) within the PDF using the cursor in the page view pane to define a new tagged region. Alternatively, they can opt for automatic detection of the regions. The automatic region detection uses a combination of two AI models. For detecting formulas, we trained a Generalized Focal Loss V2 object detector with a ResNetXt101 backbone \cite{li_generalized_2021} with the FormulaNet dataset \cite{schmitt-koopmann_formulanet_2022}. Our trained model can detect locations of mathematical formulas with an accuracy of 0.89 mAP. To detect headings, text, tables, figures, and lists, a second model was trained with the same architecture and the PubLayNet dataset \cite{zhong_publaynet_2019}, reaching an accuracy of 0.90 mAP. Each PDF operator is assigned to the predicted bounding box with the highest overlap. If two bounding boxes yield similar overlap, the bounding box with the higher score is selected. If a PDF operator does not overlap with a bounding box, it is marked as Artifact. This ensures that all PDF operators are marked. The detected regions are outlined in the page view with the region type name in the top right corner of the outline box. The color of the outline boxes indicates the type of region for an easy visual overview \cite{wolfe_guided_1989}.

To edit existing regions, the user has two options: they can select one or multiple regions and delete the region tags, which turns the corresponding PDF operators back to untagged. They can also resize an existing region by dragging one of the eight resize points. 
Information about a selected region is shown in the workspace. There, the user can check the text assigned to the region and change the region type using a drop-down menu. This kind of annotation is widely used in annotation tools such as PAWLS \cite{neumann_pawls_2021} or PDFAnno \cite{shindo_pdfanno_2018}.
In contrast to Pradhan et \textit{al}. \cite{pradhan_development_2022}, the user does not have to define the heading levels up front, which is difficult before all headings have been defined. 

\subsection{Step 2 - Reading Order}
In the second step, the user defines the reading order for each page, i.e. the logical order in which the regions defined in the previous step should be read by a screen reader. Research has shown that reading order is crucial for individuals who use screen readers and an incorrect reading order can lead to frustration \cite{lazar_what_2007, jembu_rajkumar_pdf_2020}.

\begin{figure*}
    \centering
    \includegraphics[width=\textwidth]{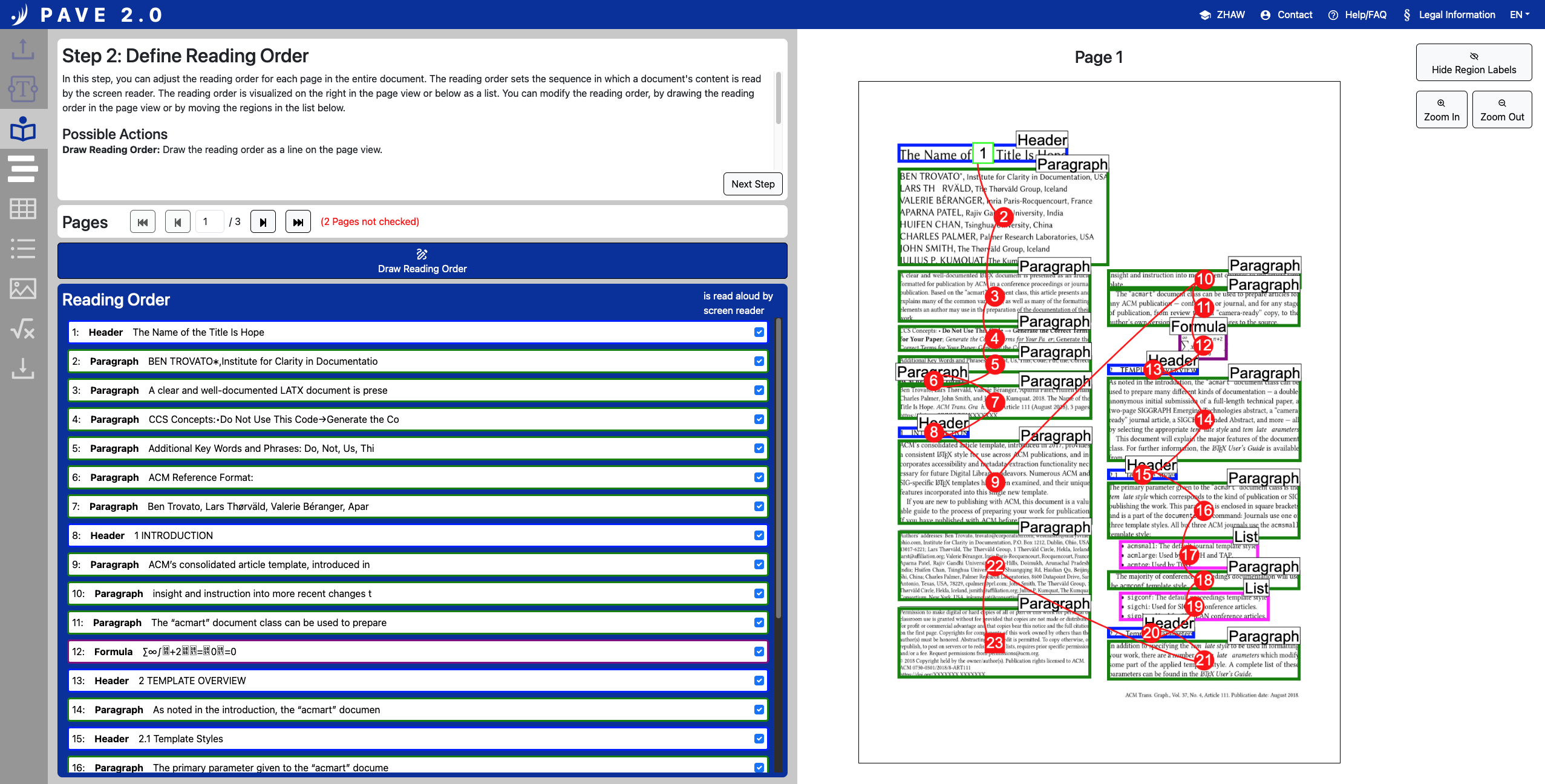}
    \caption{Screenshot of Step 2 (reading order) in PAVE 2.0.}
    \label{fig:step2}
\end{figure*}

The reading order is visualized on the page view as a directed line graph with numbers, as well as an ordered list in the workspace pane (see Figure \ref{fig:step2}). The user has two options to modify the reading order. First, they can simply draw a line into the PDF page shown the page view to define the desired reading order. If the user selects this option, all regions are first marked in red. The user can then define the reading order by drawing a curved line over the regions in the order they should be read. If the cursor crosses a region, the region text is marked green, and a green bounding box appears. If the user has not selected all regions with the drawn line graph, the skipped regions are appended to the reading order in their previous order. Alternatively, to modify the reading order only slightly, the user can move elements in the reading order list in the workspace pane. Additionally, the user can change a region into an artifact if it should not be read out at all. The color codes for the bounding boxes are the same as in Step 1.

\subsection{Step 3 - Heading Levels}
In the third step, the user defines the heading structure of all headings in the document. Headings allow individuals who use screen readers to easily navigate in a document. Research has shown that headings are the most used option to navigate websites \cite{webaim_webaim_2024}. Hence, we added an explicit step for defining the heading structure and did not integrate it with Step 1, as Pradhan et \textit{al}. did.

The heading structure of the complete document is visualized as a list in the workspace pane. There, the user can adjust each heading level using the corresponding drop-down menu. The choice of heading levels is restricted to those that adhere to a valid heading structure. The user can also allow the system to automatically detect the heading levels based on their text size (also changes existing heading levels). If an uploaded PDF already contains an invalid heading structure, the heading structure is automatically updated to a valid structure. This means that <H> tags are automatically changed to <H1>. Additionally, heading level skips, such as an <H3> heading directly following an <H1> heading, are corrected by leveling up <H3> to <H2>. In the color-coded bounding boxes the heading levels are indicated at the top right corner to simplify their validation.

\subsection{Step 4 - Tables}
In the fourth step, the user defines the structure of each table in the document. In recent years, automatic table recognition has become popular, but errors still occur \cite{kasem_deep_2024}. Therefore, explicit user interaction is still necessary to correct tables. However, various table annotating methods have been developed to create new datasets for training deep learning models. One of the most promising is the PDF Table Annotation tool \cite{frey_efficient_2015}. Inspired by their findings, we developed a table annotating method for this process step which can handle at least simple tables. Due to limitations of our back end and to first focus on the new remediation concept, tables with merged cells or tables that span multiple pages are to be addressed in future work.

\begin{figure*}
    \centering
    \includegraphics[width=\textwidth]{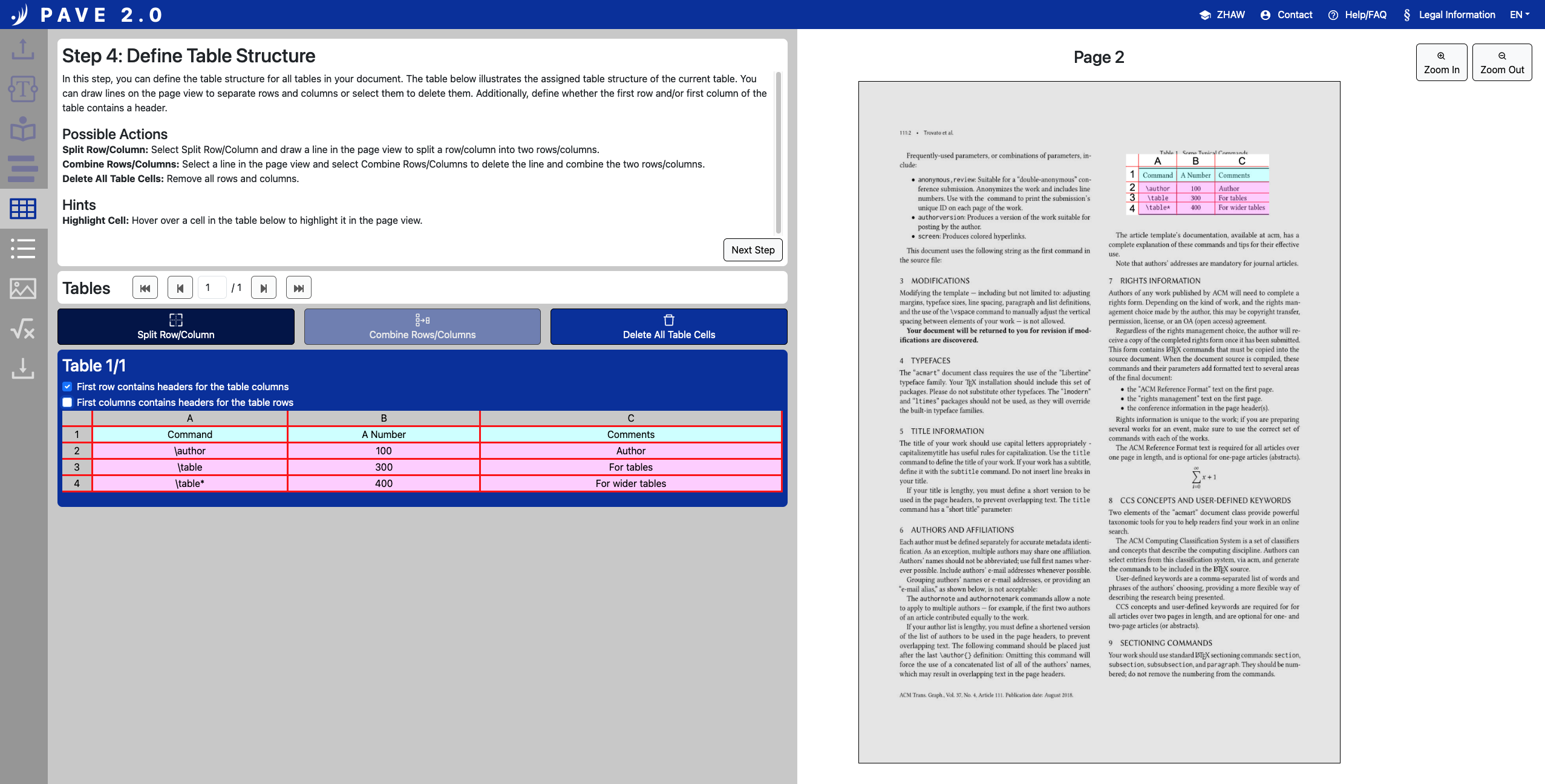}
    \caption{Screenshot of Step 4 (tables) in PAVE 2.0.}
    \label{fig:step4}
\end{figure*}

The user defines the table structure in PAVE 2.0 by drawing or deleting horizontal and vertical lines in the page view to separate rows and columns (see Figure \ref{fig:step4}). The tagged table cells are displayed in the workspace pane for reference. The user can also specify whether the first row, first column, or both are header cells. If users upload a PDF with an invalid table tag structure, the table tag structure is automatically updated to meet the UA standard.

\subsection{Step 5 - Lists}
In the fifth step, users define the structure of all lists in the document. The list interface screen is analogous to the table interface screen for consistency. To separate the list items, the user can again draw (horizontal) lines in the page view, and the list is shown on the left side in the workspace pane. The drop-down menu there can be used to create a nested list (a list within a list). As in the previous steps, the user can only create a valid list structure. Furthermore, existing invalid list structures are updated to be valid structures. For the same reasons as in the table step, defining a list spanning multiple pages is to be addressed in future work. 

\subsection{Step 6 - Figures}
In the sixth step, users can modify the alternative texts for all figures within the document. To encourage brevity a word counter is integrated into the alternative text editor, adhering to recommendations for shorter alt texts \cite{web_accessiblity_intitivate_wai_wcag_2023}. Nonetheless, research by Williams et \textit{al}. \cite{williams_toward_2022} indicates that longer alt texts are often of higher quality. To balance these findings with the complexity of scientific graphics, we added a word counter which counts down from 50 words, which typically equates to two to four sentences. Nevertheless, users can still add longer alt texts if needed (the word counter will get negative and red). Additionally, users have the option to mark a figure as decorative.

\subsection{Step 7 - Mathematical Formulas}
\begin{figure*}
    \centering
    \includegraphics[width=\textwidth]{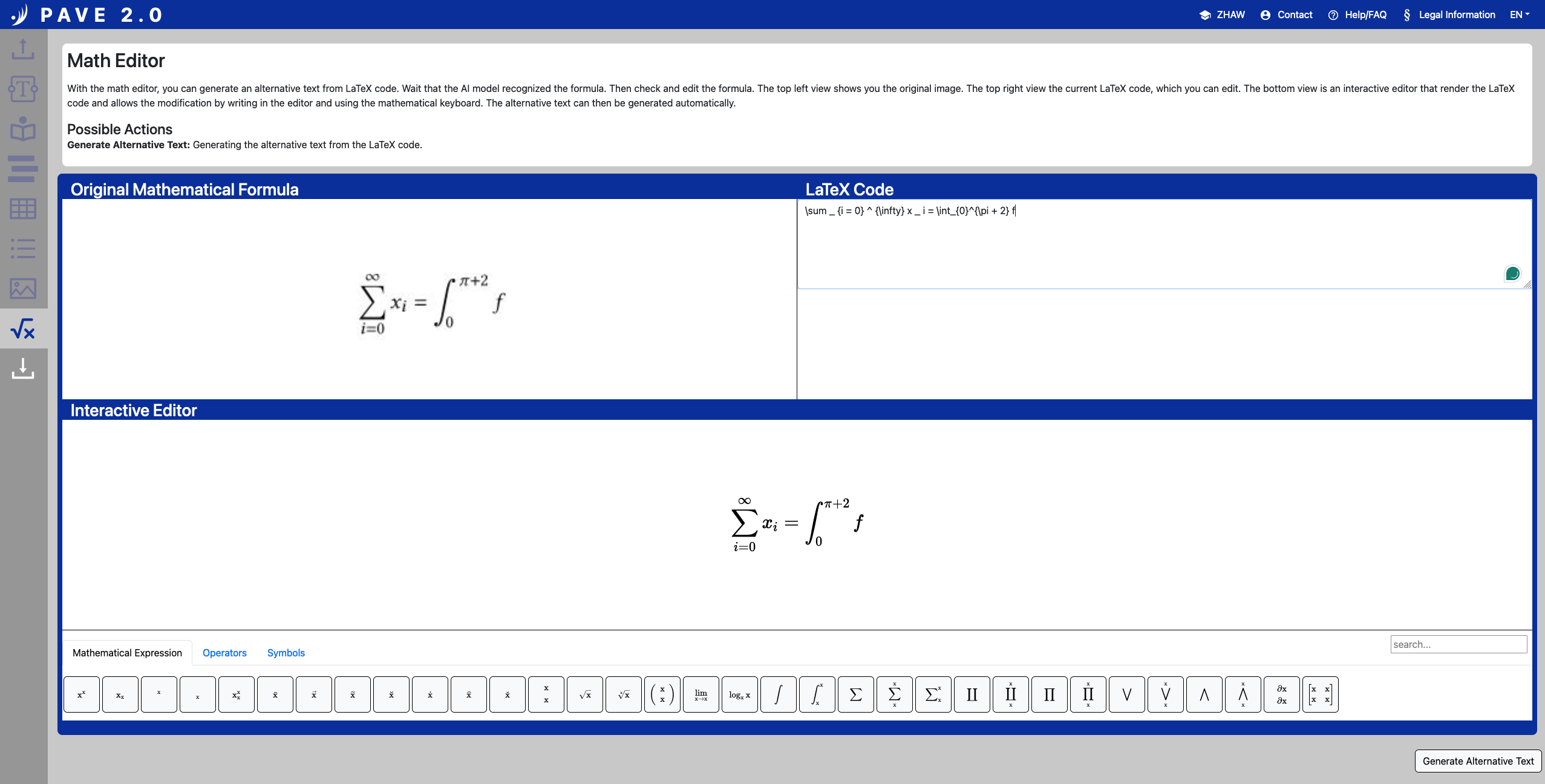}
    \caption{Screenshot of the math editor. The top left box shows the original formula from the PDF. The top right box shows the LaTeX code. At the bottom is the interactive editor with a mathematical keyboard. }
    \label{fig:mathEditor}
\end{figure*}

In step seven, users can modify the alternative texts for all mathematical formulas in the document. Alternative texts for mathematical formulas must adhere to specific rules to prevent ambiguity, which makes it challenging to write valid alternative text, even for experts \cite{fateman_how_1998}. We therefore developed a math editor (Figure \ref{fig:mathEditor}) which allows users to create valid alternative texts for formulas without knowledge of alternative text rules for formulas such as MathSpeak \cite{seewritehear_mathspeak_2021}.

After opening the math editor, an AI model based on the work of Schmitt-Koopmann et \textit{al}. \cite{schmitt-koopmann_mathnet_2024} predicts the corresponding LaTeX code. We used the same architecture and training process, but adapted the preprocessing pipeline for PDFs. The PDF is converted into a PNG image with a resolution of 600 DPI by using the pdf2image library \cite{belval_belvalpdf2image_2025}. The PDF rendering information is used to determine the position of the formula in the PDF file so that it can be cut out of the image. The AI model uses a Convolutional vision Transformer \cite{wu_cvt_2021} with 3 layers as Encoder and a Decoder Transformer with 8 heads and 4 layers as decoder. The model reaches an Edit score of 97.2\% on the im2latexv2 test dataset \cite{schmitt_koopmann_2024_11230382}, indicating an average error rate of 3 symbols per 100 symbols in an equation.

The original image of the formula in the PDF is shown in the top left subpane in the workspace pane. The LaTeX code generated by the AI model is displayed in the top right subpane. At the bottom, an interactive editor renders the LaTeX code. Users can correct the formula recognized by the AI model in two ways. First, they can modify the LaTeX code directly in the code pane. Alternatively, users unfamiliar with LaTeX can use the interactive formula editor with the mathematical keyboard at the bottom. As soon as the user has corrected the formula the math editor automatically converts the LaTeX code into an alternative text following the MathSpeak rules.

\subsection{Step 8 - Meta Information and Page Review}
\begin{figure*}
    \centering
    \includegraphics[width=\textwidth]{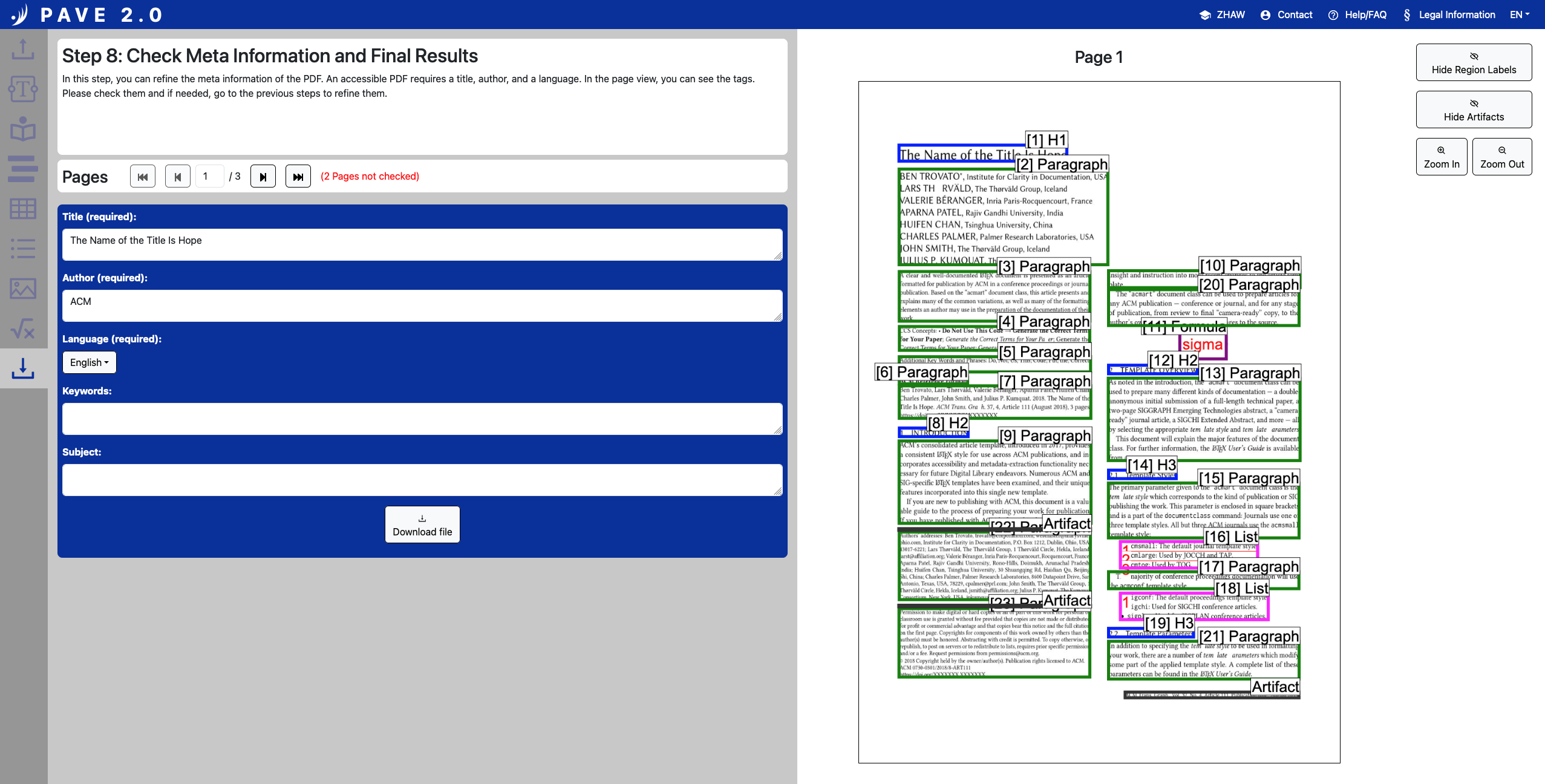}
    \caption{Screenshot of Step 8 (meta information) in PAVE 2.0.}
    \label{fig:step8}
\end{figure*}

In the last step, users can review each page of the document with an overlay of the added structure (refer to Figure \ref{fig:step8}). They can also modify the metadata information including title, author, and language. All other meta information required for the UA standard is set automatically. 

\section{User Study Design}
\label{sec:UserStudyDesign}
\subsection{Study Goal}
We aimed to answer two primary questions with this study. Firstly, how do users feel about the experience of using the guided semi-automatic PAVE 2.0 process, the interface design, ease of navigation, and overall usability of PAVE 2.0 tool? Secondly, how does the quality of tags and processing time with PAVE 2.0 compare to the most commonly used tool, Adobe Acrobat Pro?

\subsection{Participant Recruitment}
Most people are unfamiliar with PDF remediation, so we decided to conduct our study online to reach a broad audience. Furthermore, as our tool is designed for scientific documents, we specifically targeted individuals with a scientific and academic background.

The study by Pradhan et \textit{al}. \cite{pradhan_development_2022} inspired our recruitment strategy. Following their idea to maximise participant recruitment, we required only that participants be aware of PDF accessibility. To achieve a similar balance between novice and experienced participants, we aimed to recruit a study population in which half of the participants had remediated fewer than 10 PDFs (novice users) and the other half had remediated at least 10 PDFs (experienced users). We distributed the study invitation to relevant online communities, such as the WHO GATE community, and within our social and professional networks.

We recruited 19 participants for our study. One of the participants was aged between 18 and 30, nine were aged between 31 and 40, six were aged between 41 and 50, and three were aged between 51 and 60. Thirteen participants identified as female, four as male, one as non-binary, and one did not disclose their gender. Eleven participants were located in Switzerland, five were in Austria, one was in Germany, and two were in Spain.

\subsection{Study Procedure}
We compared the efficiency and effectiveness of our novel process and tool with the de facto standard PDF remediation tool Adobe Acrobat Pro. The study was conducted online via MS Teams or ZOOM depending on participants' preferences. We recorded the screen and audio to analyze the interactions and transcribe the interviews.

The study was separated into four parts.  In the first part of the study, we welcomed the participants, and collected demographic information and information about their previous PDF remediation experience. In Parts 2 and 3, the participants were asked to complete a remediation task with each of the two PDF remediation tools (one tool in Part 2 and the other tool in Part 3). To mitigate bias and learning effects, we counterbalanced the order of the tools. After each task, we conducted a semi-structured interview to gather insights into their experiences with the tool. The task is described in Section \ref{subsec:UserStudyTask}. Part 4 consisted of a structured interview with closed-ended questions. In this part, the participants were asked to compare the two tools through questions like "Which tool do you prefer to use in the future?" and "Which of the tools makes your work faster and more efficient?". 

\subsection{Task Design}
\label{subsec:UserStudyTask}
After setting up and trying out the tool, participants were asked to remediate a scientific PDF without tags. To keep the task duration manageable the PDF was a shortened 3-page version of the ACM LaTeX template with a two-column structure. The final PDF contained seventeen headings, four lists, one figure, one table, and three mathematical formulas. We created two slightly different versions of the PDFs and counterbalanced them to ensure that the order of the PDF versions did not affect the results. For the sake of ecological validity, we wanted the task to reflect real world use of PDF remediation tools, so we allowed the participants to use any additional support they wanted, such as Google.

We decided against creating detailed tutorials for each tool and provided only a brief introduction to each tool as we were also interested in whether the tools were self-explanatory and easily learnable without explicit instruction. For Adobe Acrobat Pro, we helped participants find the accessibility tools and showed them how the accessibility report works. Additionally, we showed them the help page for Adobe Acrobat's accessibility tools. For PAVE 2.0, we explained that the tool is split into eight steps, and that they could always find a description of each step in the top left corner. To gain some initial familiarity with each tool, they were allowed to try it out for up to 5 minutes with a sample PDF, similar to the user study by Pradhan et \textit{al}. To limit the duration of the study, we set a time limit of 20 minutes. If the participants were stuck for more than 3 minutes, we helped them so they could continue.

\subsection{Data Analysis}
After obtaining approval from our organization's ethical committee, we conducted the user study via MS Teams or ZOOM. The first author conducted all the interviews. During the user study, we recorded participants' audio and screens. The interviews were transcribed automatically with MS Teams/OneDrive and manually corrected to ensure a high-quality transcript. The interviews were analyzed using deductive and inductive coding. Prior to the analysis, an initial coding scheme was developed based on our research questions and the interview guide. Following Saldaña's method \cite{saldana_coding_2013}, a researcher who was not the interviewer assigned codes to relevant content during the first cycle of analysis. In a second cycle, codes were combined to form more abstract categories. The interviewer and the researcher who had assigned the initial codes executed the second cycle together. 

In addition to collecting qualitative feedback we used two quantitative metrics. First, we analyzed the screen recordings and determined the time per step in PAVE 2.0. The time duration was summed up if the participants jumped forth and back between steps. Since Adobe Acrobat has no clear workflow, we could not split the time up into steps. Second, we assessed the accessibility of the remediated PDFs created by the participants by analyzing the structure tree with Adobe Acrobat Pro and our previously defined tag accuracy criteria defined in section \ref{subsec:TagAccuracyCriteria}. The analysis of the tag accuracy was done by the interviewer for all PDFs.

\subsection{Tag Accuracy Criteria}
\label{subsec:TagAccuracyCriteria}
The PDF accessibility analyses presented in Section \ref{sec:RelatedWork} evaluate the accessibility of a PDF based mainly on metadata and machine-checkable criteria. However, these methods do not reveal other concerns, for example whether all headings are correctly tagged \cite{kumar_uncovering_2024}. Therefore, we defined thirteen manually checkable criteria to evaluate the tag accuracy of a tagged PDF using the Matterhorn Protocol \cite{pdf_association_matterhorn-protokoll_2020}, the best-practice guide of the PDF Association \cite{pdf_association_tagged_2023}, and the synthetic errors defined by Pradhan et \textit{al}. For each criterion we used the following score:
$$score = \frac{CT}{CT + WT} \cdot 100 \%$$
Correct Tag (CT) refers to the number of elements tagged accurately. Wrong Tag (WT) indicates the number of elements that should have been tagged but were not tagged, were tagged incorrectly, or should not have been tagged but were tagged. We determined the number of CTs and WTs by analyzing manually the structure tree with Adobe Acrobat Pro. We would like to emphasize that evaluating these criteria requires much experience with tagged PDFs and is only recommended for PDF accessibility experts. The criteria we used are as follows:

\begin{itemize}
    \item \textbf{All Content Tagged}: Checks if all elements in a PDF are tagged (Matterhorn checkpoint 01-005).
    \item \textbf{Reading Order}: Checks if a page is completely tagged and the reading order is not disrupted (Matterhorn checkpoint 09-001). For example, a figure appearing in a sentence is counted as an error. However, if a figure appears after Paragraph 1 instead of Paragraph 2 this is not counted as an error.
    \item \textbf{Headings Tagged}: Checks if the heading is tagged as a heading, independent of the level (Matterhorn checkpoint 14-001). Since the document title can be tagged as a paragraph or an H1, we counted both as a correct heading.
    \item \textbf{Headings Tagged + Level}: Checks if the Headings Tagged criterion is fulfilled and if the heading has the correct heading level (Matterhorn checkpoints 14-002, 14-003, 14-004, and 14-005). Similar to the criteria Headings Tagged, the document title could be tagged as a paragraph or a level 1 heading. However, depending on the title tag, the levels of the other headings must be adjusted.
    \item \textbf{Tables Tagged}: Checks if all PDF elements that are part of a table are in the table tag (Matterhorn checkpoints 15-005). 
    \item \textbf{Tables Tagged + Structure}: Checks if the Tables Tagged criterion is met and if the table structure (<TR>, <TH>, <TD>) is correct (Matterhorn checkpoints 09-002, 09-004, 15-001, and 15-002).
    \item \textbf{Lists Tagged}: Checks if all PDF elements part of a list are in the list tag (Matterhorn checkpoint 17-001). The focus is on elements that visually appear as lists. For example, a collection of bullet points or numbered items should be tagged as a list. Conversely, elements that are not visually presented as lists, such as authors of a paper, are not required to be tagged as lists. If such elements are tagged as a list, they will not be counted as errors.
    \item \textbf{Lists Tagged + Structure}: Checks if the Lists Tagged criterion is fulfilled and if the list items are separated correctly with <LI> tags (Matterhorn checkpoints 09-002 and 09-005).
    \item \textbf{Figures Tagged}: Checks if all PDF elements that are part of a figure are in the figure tag (Matterhorn checkpoints 13-001 and 13-006). If the caption tag was included in the figure, it was not counted as an error.
    \item \textbf{Figures Tagged + Alt Text}: Checks if the Figures Tagged criterion is fulfilled and if an alternative text was set (Matterhorn checkpoints 13-004 and 13-005). The quality of the alternative text is not evaluated.
    \item \textbf{Formulas Tagged}: Checks if all PDF elements that are part of an isolated formula are in the formula tag (Matterhorn checkpoint 17-001). We do not count it as an error if the formula reference was part or not of the formula tag. Formulas embedded in the text are not checked, as the distinction between embedded formulas in the text and standard text requires a detailed analysis of the text. Furthermore, it is unclear whether single symbols like $\alpha$ should be tagged as formulas with alternative text.
    \item \textbf{Formulas Tagged + Alt Text}: Checks if the Formulas Tagged criterion is fulfilled and if an alternative text was set (Matterhorn checkpoint 17-002). The quality of the alternative text is not evaluated.
    \item \textbf{Captions}: Checks if all PDF elements that are part of a figure caption or table caption are in the caption tag (Matterhorn checkpoint 13-003).
\end{itemize}

\section{Results}
\label{sec:Results}
We asked participants how often they make PDFs accessible: six novice participants had never done it before, and one novice participant had attempted it only once. 
One novice participant reported remediating PDFs approximately once a year, three experienced participants stated they did it a few times a year, one novice and four experienced participants did it approximately once a month, and one novice and two experienced participants approximately once a week.

Participants had knowledge of various tools for making PDFs accessible. The most frequently mentioned was Adobe Acrobat Pro's accessibility tools (16 participants). Participants were also aware of PAVE (8 participants) and Microsoft Office's accessibility export tools (7 participants). A minority mentioned other tools, such as Axes4 or PAC. They had used these tools to make various documents accessible, including reports, presentations, and educational materials such as exercises and handouts.

In section \ref{subsec:QuantativeResulsts} we present the tag accuracy scores, data regarding time spent, and quantitative feedback. Section \ref{subsec:Pave2Experience} summarizes the qualitative feedback on PAVE 2.0. Section \ref{subsec:AdobeExperience} summarizes the qualitative feedback on Adobe Acrobat Pro.

\subsection{Quantitative Results}
\label{subsec:QuantativeResulsts}
\subsubsection{Tag Accuracy Results}
To evaluate the tag accuracy in the remediated PDFs, we used the criteria defined in Section \ref{subsec:TagAccuracyCriteria}. Table \ref{tab:TagQuality} shows that on average experienced users remediated the PDFs 90.7\% better with PAVE 2.0 than with Adobe Acrobat Pro for all categories. Novice users remediated the PDFs 91.8\% better on average with PAVE 2.0 for all categories except the reading order. However, the difference between experienced and novice users is relatively small for both tools, averaging a 4.9 pp. difference with PAVE 2.0 and an average difference of 2.8 pp. with Adobe Acrobat Pro. Interestingly, similar to the findings of Pradhan et \textit{al}., novice users could correct the reading order better than experienced users with Adobe Acrobat Pro. The low "All Content Tagged" score of experienced users with Adobe Acrobat Pro (44.4\%) shows that only 4 of 9 PDFs were completely tagged. This indicates that the majority of the experienced users could not finish the tagging process in time. The reasons are that experienced users tagged the PDF manually or deleted larger parts of the automatically created tags.

To determine the influence of each tool's automatic tagging function, Table \ref{tab:TagQuality} also includes their respective auto-tagging scores. We want to emphasize that these auto-tagging scores depend heavily on the characteristics of the PDF itself. Hence, we want to share our anecdotal experience and insights with the auto-tagging functionalities of PAVE 2.0 and Adobe Acrobat Pro to provide a comprehensive overview. We observed that the auto-detecting of the regions in PAVE 2.0 works reliably for scientific documents, such as conference papers, but shows weaknesses with non-scientific documents. Due to the limitation of classes in the training dataset, it cannot detect captions, footnotes, and other infrequently used content types, and typically tags them as paragraphs. Due to the recognition of heading levels based on the font information, heading levels that use the same font (size and style) are not recognized correctly. The mathematical formula recognition works reliably for single-line mathematical expressions but shows weaknesses for single mathematical symbols or multi-line mathematical expressions. Adobe Acrobat Pro's auto-tagging function often fails to detect inline headings and the correct heading level. Additionally, mathematical formulas and captions are often not tagged correctly. While the list structure is usually detected very well, the list tag sometimes does not contain the complete list. Similarly, if tables are recognized correctly, the table structure is recognized often very well.

The auto-tagging results show that PAVE 2.0's auto-tagging reaches an average score of 56.9\%,  19.4 pp. higher than Adobe Acrobat Pro. Additionally, our results showed that experienced and novice users could improve the auto-tagging results by 23.2 pp. and 18.3 pp. respectively through manual correction using PAVE 2.0. In contrast, experienced and novice users could only improve the auto-tagging results by 4.5 pp. and 1.7 pp. when using Adobe Acrobat Pro.

Furthermore, the results show that users could improve the tag accuracy of the auto-tagging with PAVE 2.0 for all criteria, except for the Figures Tagged and Formulas Tagged + Alt Text criteria. The Figures Tagged score is lower because some users erroneously marked the figure as an artifact. The lower score for Formulas Tagged + Alt Text criteria occurred because several users did not reach Step 7 before time ran out. The scores for participants who reached Step 7 were 87.5\% for novice users and 100\% for experienced users. Interestingly, the caption scores for PDFs remediated by both tools are very low. These low caption scores can be explained by the fact that both Adobe Acrobat Pro and PAVE 2.0 automatically tag the captions as paragraphs. Hence, if the user is unfamiliar with the caption tag, the caption will remain a paragraph.

The reading order scores with PAVE 2.0 for PDFs from experienced and novice users are surprisingly low. The main reason for this is the typical structure of the first page of the PDF. Most participants defined the reading order from the top to the bottom, with the left column first and the right column second. However, this resulted in the ACM reference format in the bottom left corner of the first page being incorrectly read in the middle of the normal text paragraph (see Figure \ref{fig:step2}). Hence, the first page was incorrect in 15 of 19 PDFs remediated with PAVE 2.0, accounting for 78\% of errors among experienced users and 38\% of errors among novice users. In contrast, using Adobe Acrobat Pro, the first page had an incorrect reading order in 17 out of 19 PDFs, representing 25\% of experienced users' errors and 80\% of novice users' errors. This corresponds to our observation that most participants did not review or correct the automatically generated reading order.

To compare the tag accuracy scores reached in our study with current scientific papers, we calculated the tag accuracy scores of the existing tags for 10 papers from each of three popular accessibility research conferences (ASSETS 2023 \cite{ASSETS23Proceedings}, CHI 2024 \cite{CHI24Proceedings}, ICCHP 2024 \cite{miesenberger_computers_2024}). The papers were selected randomly with the help of a number generator, while the order of the papers in the proceedings determined the paper number. For ASSETS and ICCHP all 10 papers were tagged. For CHI 8 of 10 papers were tagged, as a result, we calculated the scores based on the 8 tagged PDFs. The CHI papers had on average the most pages with 14.25 without references and appendix (ASSETS: 12.2, ICCHP: 8.3). CHI papers also contained most headings per page with 3.16 (ASSETS: 3.11, ICCHP: 2.08). Figures are the most popular component of the four components (tables, lists, figures, and formulas) for all conferences (CHI: 0.61 figures per page, ICCHP: 0.52, and ASSETS: 0.4). Formulas instead are the least popular component (CHI: 0.11 formulas per page, ICCHP: 0.04, and ASSETS: 0.01). Interestingly, ASSETS papers contained more lists per page (0.21) than CHI papers (0.17) and ICCHP papers (0.10), while for the other component types CHI papers contained the most components per page. The component counts in our sample differ slightly from those reported by Menzies et al. \cite{menzies_author_2022} for 330 ASSETS papers between 2011 and 2020. On average, they observed 2.0 tables per paper (1.7 in our sample), 9.2 figures per paper (5.0 in our sample), and 0.3 formulas per paper (0.1 in our sample).

We observed that the tag accuracy and tagging strategy vary greatly from paper to paper for all conferences. For instance, some papers tagged inline headings (level 3 in the ACM template) as headings, while others tagged them as paragraph. We also observed that in some papers, the caption tag is nested within the figure tag, and in others, it appears after the figure tag. This inconsistent tagging strategy indicates that the papers were tagged with different (interpretations) or no guidance. The average score indicates that the conference papers are slightly more accessible compared to the results with Adobe Acrobat Pro, but clearly lower than with PAVE 2.0.
We observed that ASSETS and CHI papers never tagged the "Check for updates" button (figure with a link), while it was tagged in 7 of 10 of the ICCHP PDFs. Failing to tag the "Check for updates" button is also the reason for the 0 scores in the criteria "All Content Tagged" for the ASSETS and CHI papers. Furthermore, the evaluation shows the great value ASSETS and CHI put on tagging figures with alt text, while the ICCHP papers never included alternative text. A reason for the high reading order score of the ICCHP PDFs could be the simpler one-column layout, compared to the two-column layout of ASSETS and CHI.
\begin{table*}[ht]
    \centering
    \begin{tabular}{l||c|c|c||c|c|c||c|c|c}
         & \multicolumn{3}{c||}{\textbf{Adobe Acrobat Pro}}&\multicolumn{3}{c||}{\textbf{PAVE 2.0}} & \multicolumn{3}{c}{\textbf{Conferences}}\\
         Criteria & Exp. & Novice & Auto & Exp. & Novice & Auto & ASSETS & CHI & ICCHP\\
         &  [\%] &  [\%] & [\%] & [\%] & [\%] & [\%] & [\%] & [\%] & [\%]\\
         \hline
         \hline
         All Content Tagged & 44.4 & 90.0 & 100.0 & \textbf{100.0} & \textbf{100.0} & 100.0 & 0.0 (10) & 0.0 (8) & 70.0 (10)\\
         \hline
         Reading Order & 37.0 & \textbf{66.7} & 66.7 & \textbf{66.7} & 63.3 & 0.0 & 59.8 (122) & 56.1 (114) & 91.6 (83)\\
         \hline
         Headings Tagged & 83.7 & 75.3 & 70.6 & \textbf{100.0} & 98.2 & 94.1 & 71.8 (379) & 72.8 (360) & 47.4 (173)\\
         + Level & 76.5 & 54.1 & 0.0 & 79.7 & \textbf{86.5} & 70.6 & 71.0 (379) & 72.5 (360) & 42.8 (173)\\
         \hline
         Tables Tagged & 33.3 & 10.0 & 0.0 & \textbf{100.0} & \textbf{100.0} & 100.0 & 66.7 (17) & 73.7 (19) & 43.8 (16)\\
         + Structure & 0.0 & 10.0 & 0.0 & \textbf{77.8} & 70.0 & 0.0 & 29.4 (17) & 31.6 (19) & 0.0 (16)\\
         \hline
         Lists Tagged & 52.8 & 60.0 & 75.0 & \textbf{100.0} & 92.5 & 75.0 & 57.7 (26) & 31.6 (19) & 75.0 (8)\\
         + Structure & 50.0 & 60.0 & 75.0 & \textbf{77.8} & 72.5 & 0.0 & 57.7 (26) & 31.6 (19) & 75.0 (8)\\
         \hline
         Figures Tagged & 55.6 & 40.0 & 100.0 & \textbf{77.8} & 70.0 & 100.0 & 90.0 (50) & 92.9 (70) & 100.0 (43)\\
         + Alt Text & \textbf{55.6} & 30.0 & 0.0 & \textbf{55.6} & 50.0 & 0.0 & 90.0 (50) & 92.9 (70) & 0.0 (43)\\
         \hline
         Formulas Tagged & 37.0 & 13.3 & 0.0 & \textbf{100.0} & \textbf{100.0} & 100.0 & 0.0 (1) & 61.5 (13) & 0.0 (3)\\
         + Alt Text & 14.8 & 0.0 & 0.0 & \textbf{77.8} & 70.0 & 100.0 & 0.0 (1) & 0.0 (13) & 0.0 (3)\\
         \hline
         Captions Tagged & 5.6 & 0.0 & 0.0 & \textbf{27.8} & 5.0 & 0.0 & 3.0 (67) & 10.2 (88) & 61.0 (59)\\
         \hline
         \hline
         Average Score & 42.0 & 39.2 & 37.5 & \textbf{80.1} & 75.2 & 56.9 & 45.9 & 48.3 & 46.7\\
    \end{tabular}
    \caption{Overview of the tag accuracy score of the remediated PDFs by novice, experienced users or only using the auto tagging function. To compare the results, we evaluate ten randomly selected papers of each conference (Assets 2023, CHI 2024, and ICCHP 2024) additionally. The values are percentages of elements fulfilling the criteria. The brackets indicate the total number of elements related to the criteria (CT + WT).}
    \label{tab:TagQuality}
\end{table*}

\subsubsection{Time Performance}
\label{subsubsec:TimePerformance}
Table \ref{tab:PaveTimes} shows the average time novice and experienced participants spent using each tool, as well as the number of participants who reached each step and the number of instructions required with PAVE 2.0. We have included the timing reported by Pradhan et \textit{al}. for the Ally tool, but it is important to note that the task they performed involved correcting errors in PDFs that were already tagged, which is a different task than tagging an untagged PDF.

Interestingly, novice users were faster and needed fewer instructions than experienced users with PAVE 2.0. Both novice and experienced users spent most of their time on the first two steps of PAVE 2.0. Novice users were quicker with PAVE 2.0, while experienced users spent more time with PAVE 2.0 than with Adobe Acrobat Pro. Out of the 19 participants, five finished before the 20-minute time limit with Adobe Acrobat Pro. However, all five participants mentioned that the PDF was probably not completely accessible, but they did not know how to fix the remaining issues. One participant had to give up because Adobe Acrobat Pro stopped working multiple times. Novice users spent 13 minutes and 21 seconds for the first four steps, which is similar to previous findings regarding the four subtasks of the Ally prototype (the Ally prototype only has these four subtasks in its remediation process). However, experienced users spent 15 minutes and 3 seconds for the first four steps, which is around 2 minutes longer than with the Ally tool.

The timings further reveal that optimizing steps 1 and 2 with better AI models or improved user interfaces holds the greatest potential for accelerating the remediation process. Nevertheless, automating the other steps can influence the PDF remediation speed similarly depending on the used documents. For example, figures in CHI papers are very frequent (see Table \ref{tab:TagQuality}). Hence, we believe an improved figure step with specialized AI models for scientific graphics and images could reduce the burden of the PDF remediation process of CHI papers significantly. 

\begin{figure}
    \centering
    \includegraphics[width=.5\textwidth]{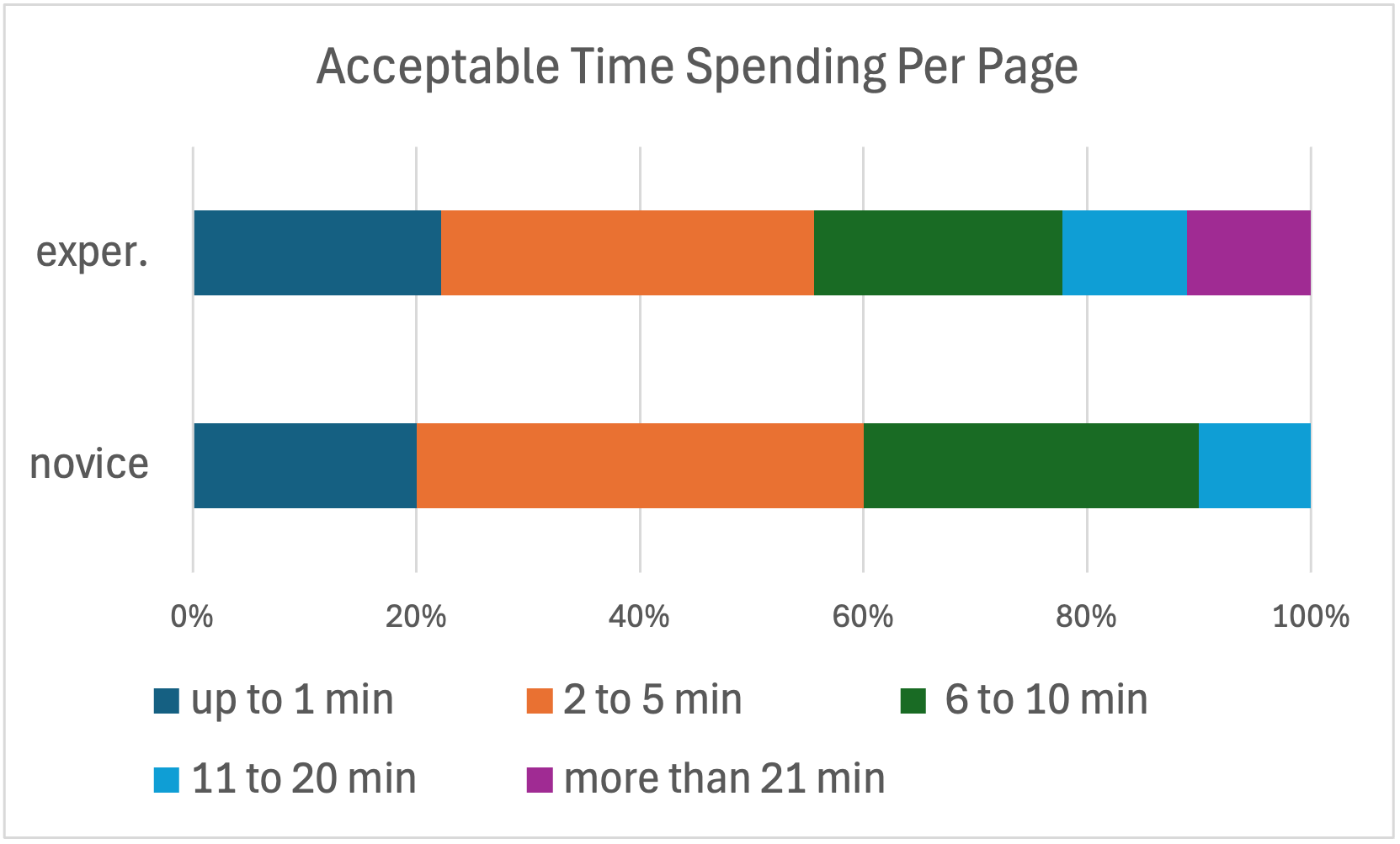}
    \caption{Answers of experienced and novice participants, about how much time they would be willing to spend per page to make a PDF accessible.}
    \label{fig:TimeSpendingRemediation}
\end{figure}

We asked participants how much time they would be willing to spend per page to remediate a PDF and categorized the responses into five ranges. As shown in Figure \ref{fig:TimeSpendingRemediation},  around 80\% of the novice and experienced users stated that 2 to 5 minutes per page would be acceptable. Novice users spent on average 5 minutes and 48 seconds per page, while experienced users spent 6 minutes and 35 seconds in our study with PAVE 2.0. This means that about 40\% of participant would already have fallen into the desired 2-5 minutes per page range in upon their first use of PAVE 2.0.

\begin{table*}[ht]
    \centering
    \begin{tabular}{c||c|c|c|c|c||c|c|c|c|c}
    & \multicolumn{5}{c||}{\textbf{Novice Users}} & \multicolumn{5}{c}{\textbf{Experienced Users}}\\
    Step & Time & Done & Instr. & Adobe & Ally  & Time & Done & Instr. & Adobe & Ally \\
    & [m:ss] & [n] & [n] & [m:ss] & [m:ss] & [m:ss] & [n] & [n] & [m:ss] & [m:ss]\\
    \hline
    \hline
    1-Regions & 3:01 & 10 & 1 & - & 4:31 & 6:25 & 9 & 1 & - & 4:46\\
    2-Reading Ord. & 6:48 & 10 & 5 & - & 3:23 & 4:36 & 9 & 4 & - & 4:10\\
    3-Heading Str. & 1:37 & 10 & 0 & - & 1:39 & 1:23 & 8 & 1 & - & 1:15\\
    4-Tables & 1:55 & 10 & 2 & - & 3:46 & 2:39 & 8 & 3 & - & 3:07\\
    5-Lists & 1:28 & 9 & 0 & - & - & 1:36 & 8 & 1 & - & - \\
    6-Figures & 0:25 & 8 & 0 & - & - & 0:52 & 7 & 0 & - & - \\
    7-Formulas & 1:12 & 8 & 0 & - & - & 1:33 & 7 & 1 & - & - \\
    8-Meta Inf. & 0:59 & 7 & 0 & - & - & 0:39 & 6 & 0 & - & - \\
    \hline
    \hline
    total & 17:25 & - & 8 & 17:30 & 13:16 & 19:44 & - & 11 & 18:34 & 13:18\\
\end{tabular}
    \caption{Overview of the average time spent by the participants for each step, split by novice and experienced users. Additionally, it shows how many participants have done a step and how many participants required instructions. Additionally, it shows the timings observed with the four steps of the Ally tool.}
    \label{tab:PaveTimes}
\end{table*}

\subsubsection{Quantitative Feedback}
\label{subsubsec:QuantitativeFeedback}
\begin{figure*}
    \centering
    \begin{subfigure}{0.4\textwidth}
         \centering
         \includegraphics[width=\textwidth]{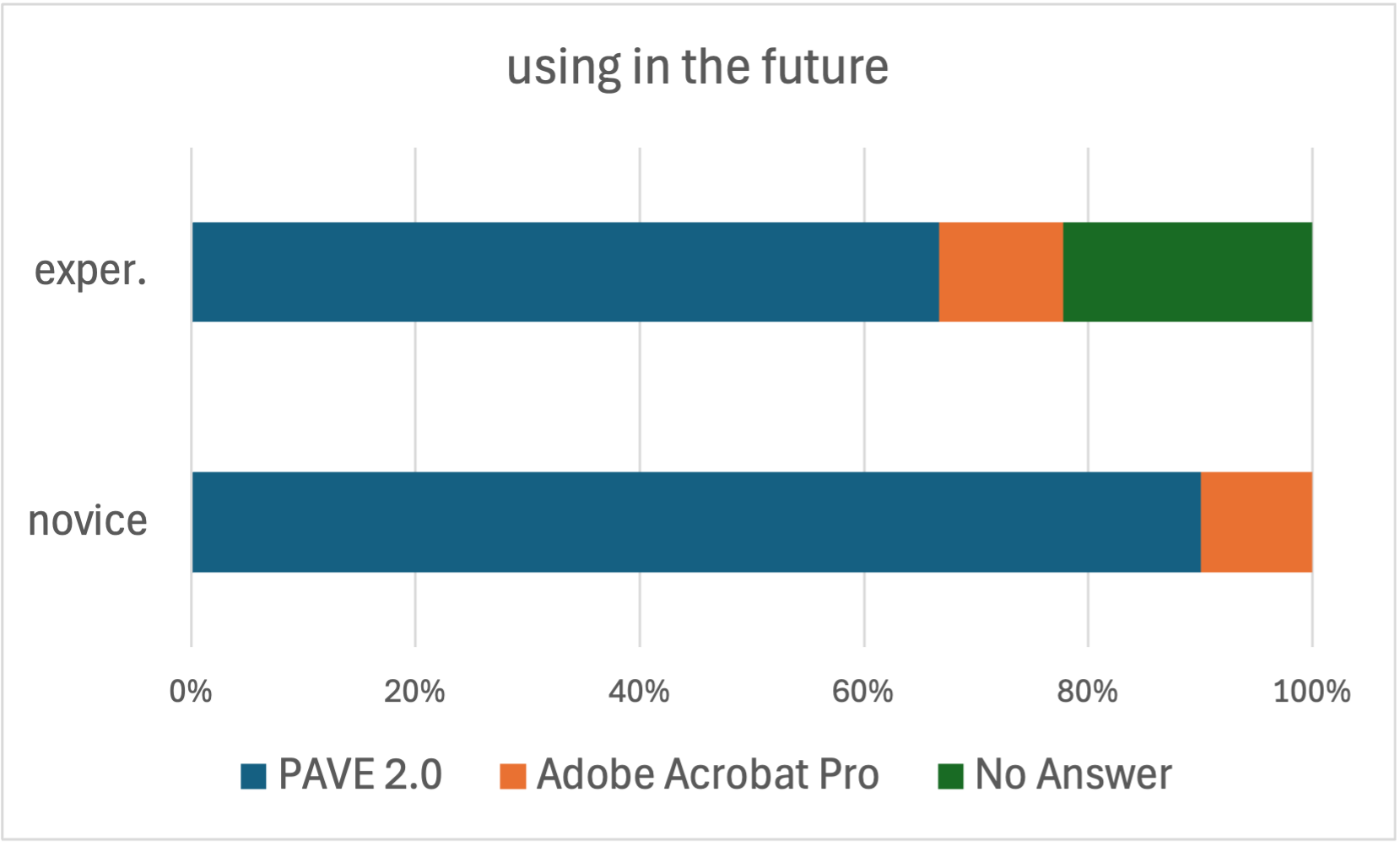}
         \caption{}
         \label{fig:usingInTheFuture}
     \end{subfigure}
     \begin{subfigure}{0.4\textwidth}
         \centering
         \includegraphics[width=\textwidth]{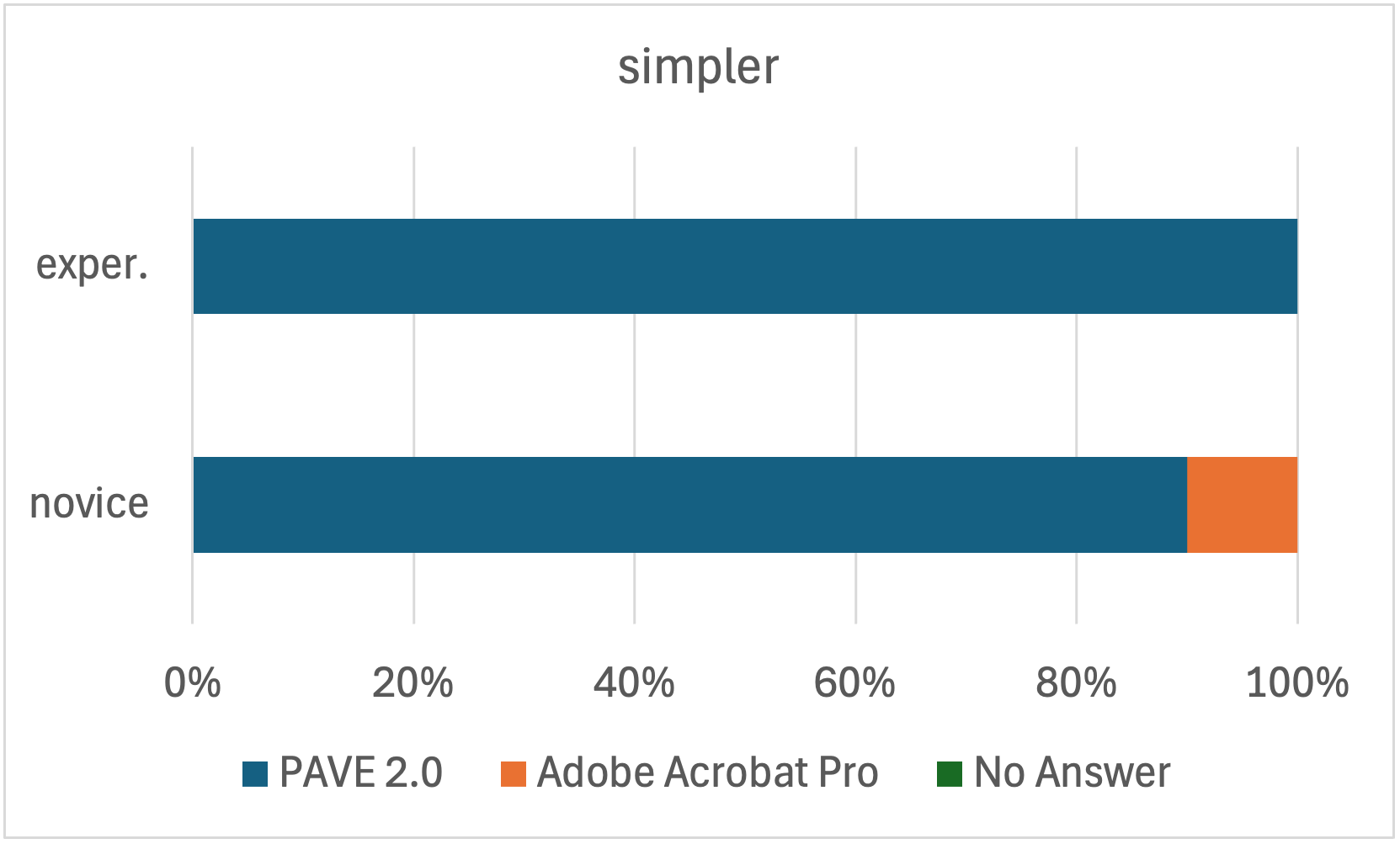}
         \caption{}
         \label{fig:Simpler}
     \end{subfigure}
     \begin{subfigure}{0.4\textwidth}
         \centering
         \includegraphics[width=\textwidth]{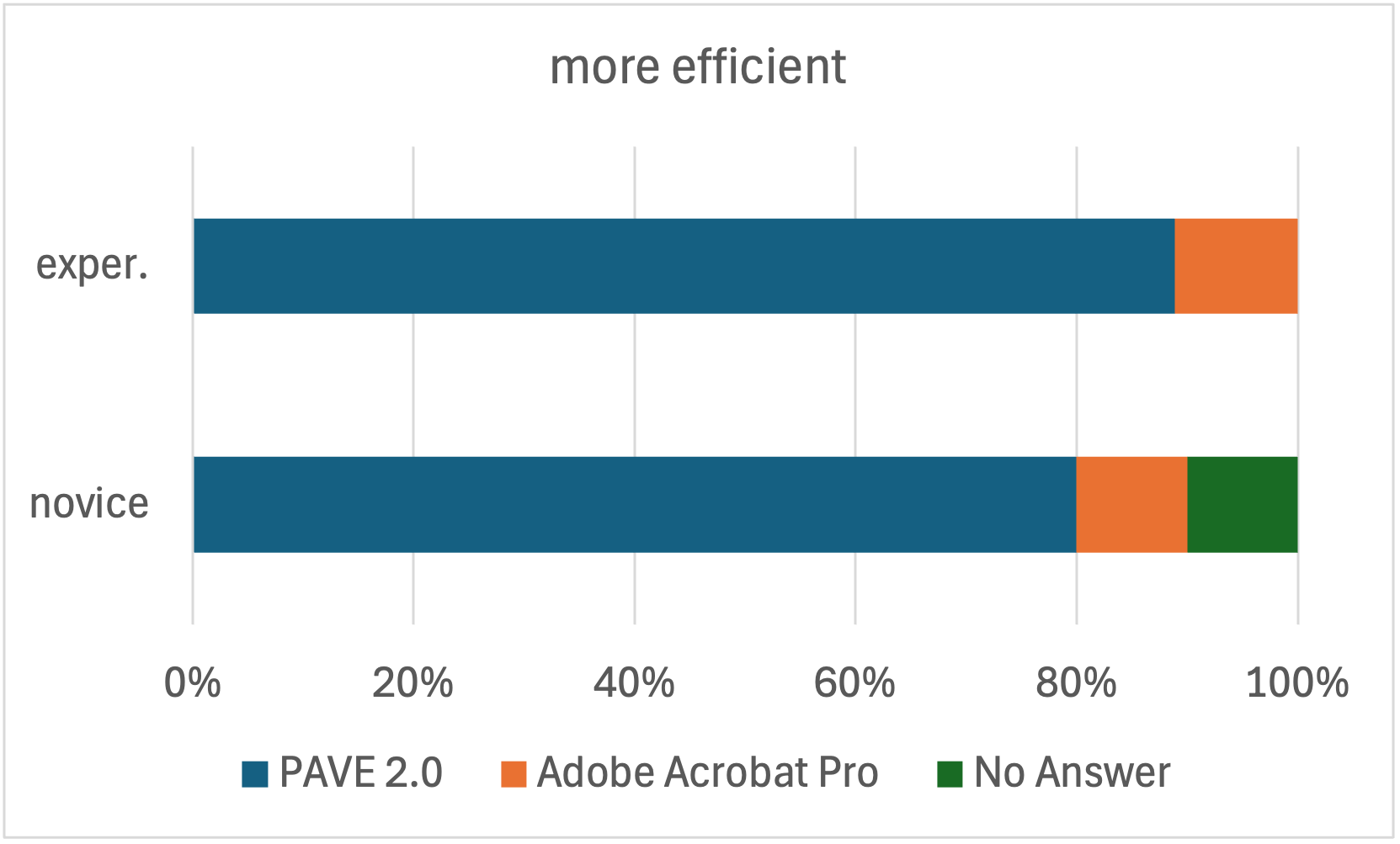}
         \caption{}
         \label{fig:moreEfficient}
     \end{subfigure}
     \begin{subfigure}{0.4\textwidth}
         \centering
         \includegraphics[width=\textwidth]{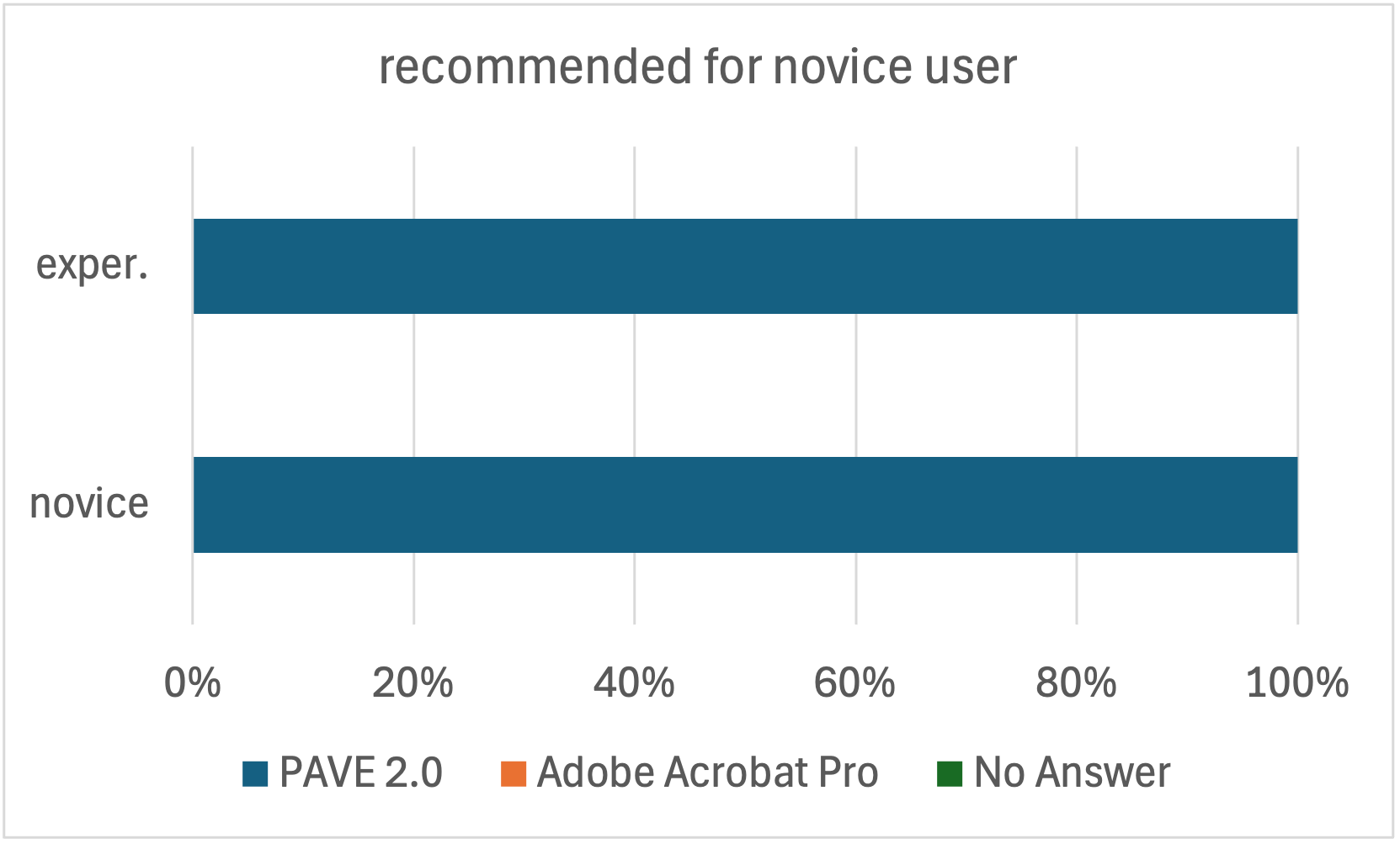}
         \caption{}
         \label{fig:recommendedForNoviceUsers}
     \end{subfigure}
    \caption{Quantitative feedback of novice and experienced users of the following four questions: a) Which tool do you prefer to use in the future? b) Which of the tools is easier to use? c) Which of the tools makes your work faster and more efficient? d) If somebody has no experience with PDF remediation. Which of the tools would you recommend?}
    \label{fig:ComparingQuestions}
\end{figure*}

We asked participants to compare PAVE 2.0 and Adobe Acrobat across various aspects through four specific questions (see Figure \ref{fig:ComparingQuestions}). When asked which tool they would prefer to use in the future, fifteen participants indicated a preference for PAVE 2.0. Two participants stated their preference would depend on the specific PDF, alternating between Adobe Acrobat Pro and PAVE 2.0. The last two participants stated they would prefer to use Adobe Acrobat Pro in the future. One of them expressed uncertainty about PAVE 2.0's ability to handle large PDFs, and the other desired an offline tool, which is simpler compared to Adobe Acrobat Pro.

Regarding ease of use, eighteen participants found PAVE 2.0 simpler, while one participant preferred the usability of Adobe Acrobat Pro. Sixteen participants responded that they believed PAVE 2.0 could enhance their current or future work with accessible PDFs. However, two participants were unsure, citing some limitations of PAVE 2.0, and one person felt that they were more efficient using Adobe Acrobat Pro. Despite these mixed preferences, all participants agreed they would recommend PAVE 2.0 to novice users seeking to make a PDF accessible.

\subsection{PAVE 2.0 User Experiences}
\label{subsec:Pave2Experience}
Most participants started Step 1 with the "Detecting Regions" function and reviewed the results. Three participants mentioned it would be beneficial if the automatic detection of the regions would run without their needing to click on the button (however, this would delete already tagged regions in the PDF). Two participants needed clarification on how to adjust the size of regions or merge them. Additionally, we observed that six participants were uncertain about the appropriate region types. One participant mentioned that they would like to have a selection guide for the region-type selection. For instance, they would appreciate knowing the possible region types for a footer. This issue of being unaware of the possible region types could also explain why fifteen participants did not label any captions as captions. Three participants mentioned that they appreciated the color-coded region types. Nevertheless, only a few participants noticed all region-type errors made by the AI model in this step, e.g., that a heading was labeled as a list. Therefore, fourteen participants had to return to Step 1 to make corrections later on. Two participants mentioned that they would like to have an undo function in this step as well as for later steps.

We observed that defining the reading order (Step 2) was the most time-consuming step. Most of the participants began by changing the reading order via the list, which is the more time-consuming process. However, around half of the participants discovered the "Draw Reading Order" function on their own. For the remaining nine participants, we had to provide a hint for them to detect the functionality. Nonetheless, after discovering the "Draw Reading Order" function, all participants found it useful. Four participants pointed out, however, that the feature requires some training, and that the detection of the boxes could be improved. Most participants appreciated the visualization of the reading order with a line graph. However, one participant mentioned that the lines and numbers obscured some parts of the text which made the review of the reading order more challenging. Another participant stated that they would appreciate if red and green color combinations were avoided and the overall color contrast was increased, on account of the challenges they present for colorblindness.

Organizing the heading levels (Step 3) was straightforward for most participants. Only two participants used the automatic detection of the heading levels; the other participants did everything manually. Two participants were confused that they could not navigate through the pages with a page navigation bar and had to click on a heading to jump to the correct page. One participant said they would like to create an <H> tag or <Title> tag for the paper's title.

Step 4 (tables) was the first object-specific step. As with the "Draw Reading Order" function, most participants did not understand how to draw the table at first. Thirteen participants figured out how to do so themselves, while five participants required instruction. Nevertheless, eight participants mentioned that they liked this step and said it was a simple solution for tables. Two participants mentioned the issue of irregular tables and tables on multiple pages, which cannot easily be addressed with this approach. One participant stated they would have preferred if the possible options for columns and rows had been highlighted.

The list remediation step (Step 5) is similar to the one for tables. As a result, participants were already familiar with the drawing functionality and completed this step with four lists much more quickly than the table step with one table. As with the table step, two participants mentioned that having a solution for lists over multiple pages and two-column lists would be beneficial.

Step 6 (alternative text for figures) was straightforward for all participants. However, we observed that some participants were unsure of what to write as alternative text. Providing more detailed instructions and examples of alternative texts, similar to the SIGACCESS Describing Figures guidelines \cite{trewin_describing_2019}, would likely be helpful for some users. Furthermore, five participants wrote an alternative text but also marked the image as decorative. Two participants mentioned that they would like to have an automatically generated alternative text option.

At Step 7 (mathematical formulas), most participants mentioned that they usually do not work with mathematical formulas. Fourteen participants used the "Math Editor" to create alternative text for formulas. One participant needed assistance to find the math editor. Two participants mentioned that one formula looked wrong, but they did not know how to correct it because they did not understand it. One participant was unsatisfied with the automatically generated alternative text and changed it manually. However, five participants mentioned that they liked that such a solution for mathematical formulas was provided.

The meta-information step (Step 8) was straightforward for all participants. We observed that most participants filled out all meta information fields, even the optional fields. Additionally, we observed that only two participants used the visualization of the accessible PDF to check the result. One participant mentioned it would be beneficial to have the option to copy text, such as the author list, from the page view. Another participant mentioned that an HTML view to check the accessibility of the resulting PDF would be valuable.

Overall, participants expressed that they would like to be able to modify the application's layout, such as changing the size of the workspace pane, page view pane, instruction box, or font size. Some also stated that the application should also be accessible for low vision or blind users. Additionally, six participants wanted to the ability to skip steps, e.g., go from Step 2 directly to Step 5.

The workflow was generally clear for all participants. However, some participants faced difficulties in discovering and using all the functionalities offered by the tool. On a scale of 1 to 5, with 1 being not satisfied and 5 being completely satisfied, the participants rated their satisfaction with the tool as 4.3 on average. Out of the 19 total participants, 17 expressed a desire to use the tool in the future if they needed to remediate a PDF. The other 2 participants would consider using the tool, but wanted to test it further before making a final decision.

\subsection{Adobe Acrobat Pro User Experiences}
\label{subsec:AdobeExperience}
Adobe Acrobat Pro offers a variety of tools and options to enhance PDF accessibility. Tags, for instance, can be created using the "Automatically Tag PDF" function, the "Fix Reading Order" tool, or manually within the "Accessibility Tags" panel. A commonly recommended first step is to use the "Check for accessibility", which generates an accessibility report. Right-clicking on an issue in the report allows the user to view an explanation on Adobe's help page, and in some cases, the user can select "Fix". Some issues can be resolved automatically by clicking "Fix", while other issues require some extra user input. However, a clear workflow which the users should follow does not exist in Adobe Acrobat Pro.

Fourteen participants used the Adobe accessibility report as a guide after we had introduced it to them. Nine participants reported that they liked the report and it gave them some guidance. However, they faced difficulties when an accessibility issue could not be resolved automatically. Two participants even reported that they were confused by the automatic fixing of the issues because they did not understand what had changed. One reason was because they did not find sufficient information about the automatic correction on the help page of Adobe Acrobat Pro. One participant recommended integrating the help page into the tool. Another participant was not satisfied with the German translation of the help page. Three participants reported that they liked the reading order tool. However, two participants were disappointed that lists could not be selected. Two participants did not like that the grey boxes overlapped leaving them unsure about whether they had selected the correct text.

We noticed that most of the participants relied on the automatic tagging function, but only a few reviewed the resulting tags. Two participants even mentioned that they were unsure how to modify the PDF structure tree. Consequently, captions and mathematical formulas were often left untagged. Additionally, ten participants did not recognize that one of the four lists was incorrectly tagged and that the table was tagged as a paragraph by the automatic tagging. However, only one of the participants did not add an alternative text for the tagged image.

Fourteen participants stated that they found the tool complicated or unintuitive. Nevertheless, sixteen participants reported they would use the tool again, while three stated they would not. However, nine of these sixteen participants mentioned that they would only use it again because they felt they had no other option. On average, the participants' average satisfaction with the tool was 2.9 on a scale of 1 (not satisfied) to 5 (completely satisfied). 

\section{Discussion}
\label{sec:Discussion}
Similar to previous studies \cite{bigham_uninteresting_2016, jembu_rajkumar_pdf_2020, pradhan_development_2022}, we found that the weaknesses of Adobe Acrobat Pro are the unclear workflow and the lack of intuitiveness of the user interface. Additionally, we observed that most users tend to focus on fixing all accessibility issues in Adobe's accessibility reports, but they often overlook checking the accuracy of the tags. Consequently, most participants did not rectify the tagging errors of Adobe's automatic tagging. As a result, the table, captions, some headings, parts of lists, and mathematical formulas were often tagged as paragraphs. On the other hand, most participants attempted to fix the heading structure nesting issue since the report highlighted that. This indicates, similar to the findings of Kumar et \textit{al}. \cite{kumar_uncovering_2024}, that accessibility checks are helpful, but they can also provide the user with a false impression of the accessibility of a PDF.

In contrast, PAVE 2.0 guides the user through the PDF remediation process in eight steps without an accessibility report. Our study showed that separating the process into smaller steps was beneficial for two reasons: First, the users easily understood the workflow and did not get lost in the process. Second, the steps helped the user to focus on the critical tasks first. With PAVE 2.0, the user must correct the regions, the reading order, and the heading structure before continuing with the other steps. As a result, the overall structure and the important heading structure for navigation are more likely to be completed even if the user stops the remediation process early (or reached the time limit in our study).

The quantitative feedback revealed that around 20\% of participants would be willing to spend only up to 1 minute per page making PDFs accessible, necessitating a highly automated system. We believe that this wish for automation is also due to authors being aware of accessibility issues, but not knowing how to fix them. However, this reliance on automation comes with its own challenges, particularly when users do not verify the accuracy of automated outputs. This situation reveals a tension in the integration of AI into accessibility tools. This tension hinges on the balance between user trust in automated solutions and the critical need for such solutions to encourage, or perhaps even require, user verification. The insights gained suggest a necessity for designing these tools in a way that not only offers automation but also embeds mechanisms that ensure users are actively involved in the verification process. This approach could mitigate the risks associated with unchecked trust in automation, thereby enhancing the overall utility and effectiveness of accessibility tools. Our analysis, as shown in Table \ref{tab:TagQuality}, indicates that our developed user interfaces allow both novice and experienced users to detect and correct such AI errors easily. Nevertheless, user interfaces for currently non-AI-supported steps may need further adaptation to maintain effectiveness and efficiency.

\subsection{Recommendations}

We want to highlight four key points for further improvement:

First, selecting the correct region type is challenging for both novice and experienced users, as also observed by Pradhan et \textit{al}. Implementing a selection assistance could save valuable time, increase users' confidence in their choices, and increase tag accuracy. 

Second, the drawing features for defining the reading order, tables, and lists were difficult for some users to locate and use, echoing Pradhan et \textit{al}.'s findings. This difficulty arises because these functions were unexpected for most users and require some training. Despite this, the drawing option proved to be effective among most users, as indicated by the tag accuracy results. We assume an interactive tutorial or a toast component could allow users to understand and learn the interaction faster.

Third, providing more background information about accessible PDFs and the reasons of limiting certain options could enhance user confidence and reduce confusion. For instance, two participants were confused at Step 3 because our process presented only one option that complied with accessibility standards. Consequently, they could not use <P>, <H>, or <Title> tags for the heading structure or skip heading levels (e.g., jumping from <H1> to <H3>). We believe that these multiple options permitted by current accessibility standards contribute to confusion and unnecessary complexity. Therefore, a clearer, less ambiguous PDF accessibility standard would be advantageous. 

Fourth, there should be a more interactive approach for screen reader user for mathematical formulas in PDFs. We do not consider adding alternative text for formulas as a good solution, particularly for more complex formulas. Instead, we suggest an interactive solution, similar to websites \cite{cervone_adaptable_2019}, where individuals who use screen readers can navigate and explore different parts of the formula. With our method, we could create such a tree structure instead of an alternative text, as we have the LaTeX code for the formula. Nevertheless, a novel PDF standard for mathematical formulas and add-ons for screen readers must be developed.
We would like to note that the PDF/UA-2 standard, which appeared after the study, proposes the use of MathML in PDFs. This will allow navigating and personalizing the reading of mathematical formulas as suggested above. However, current screen readers (e.g. JAWS, NVDA, and VoiceOver) have not implemented these features so far.

\section{Limitations and Future Work}
The user study participants all had a scientific or academic background and were interested in accessible PDFs, which limits the generalization of the answers. Further limitations that we acknowledge are due to the experimental nature of the study. Particularly the short duration during which participants used the tools means that people's experience and performance in the real world might differ. Additionally, the interview included speculative questions. These responses might not reflect actual user behavior or preferences in real-world use. Furthermore, the use of one PDF (with two versions) limits the generalization of the results, especially the auto-tagging results. Future work should also investigate how the tag accuracy score, which focuses on technical conformance, relates to the reading experience of screen reader users.

It is also worth noting the technical limitations of our prototype. Our prototype currently only supports documents that are not overly complex, e.g., not containing tables that span multiple pages, irregular tables or multi-column lists. Additionally, the user currently does not have the option to tag footnotes or links. These special cases should be addressed in future work. Additionally, the PDF parsing library utilized cannot reliably parse all PDFs, leading to potential parsing errors.

In addition to improving the step-by-step PDF remediation process, future work should also explore how this method can be effectively integrated into the publishing workflows of conferences and journals. This includes clarifying the responsibility of who is responsible for what. Research \cite{jembu_rajkumar_pdf_2020} showed that authors feel it is the publisher's responsibility to make PDFs accessible. However, making a PDF accessible requires knowledge about the content, which would allow authors to make PDFs accessible more efficiently compared to publishers. Similar to the process of Menzies et \textit{al}. \cite{menzies_author_2022}, we think authors should make their documents accessible, while publishers shall provide tools and guidelines, and ensure the accessibility quality. A seamless integration into the publishing workflow with accessibility checks and clear responsibilities will be crucial for ensuring that accessible PDFs become a standard in academic publishing.

\section{Conclusion}
\label{sec:Conclusion}
Previous research has mainly focused on analyzing the inaccessibility of PDFs and developed accessibility methods for specific challenges. Our work builds upon this and introduces a novel PDF remediation process which reduces the knowledge necessary to remediate PDFs and, as a result, makes the method suitable for novice and experienced users. We implemented a prototype and compared it with today's de-facto standard accessibility remediation tool, Adobe Acrobat Pro, in a user study with nineteen participants. Our study demonstrated that our step-by-step process enables both novice and experienced users to remediate a PDF with around 90\% higher tag accuracy compared to Adobe Acrobat Pro, requiring minimal expert knowledge and time from authors. Two participants even reported finding our PDF remediation process fun and enjoyable. Furthermore, we developed a math editor, allowing straightforward creation and modification of a mathematical formula's alternative text, even for novice users. The generated annotations would even allow the creation of more accessible representations and interaction possibilities with a formula, e.g., a tree structure. Additionally, we presented thirteen criteria and a score to manually evaluate the tag accuracy in a PDF, addressing aspects that are not fully covered by existing accessibility checkers. 

As previous \cite{wang_improving_2021, nganji_assessment_2018, darvishy_state_2023, pierres_pdf_2024} and our analysis of the accessibility of scientific PDFs revealed, are most scientific PDFs still inaccessible. We believe integrating our step-by-step process into a conference publishing workflow could improve PDF accessibility significantly and should be the next step. We estimate it would require authors to spend approximately an additional hour when submitting a 10-page paper. However, authors would no longer need specialized PDF remediation software, and the tagging structure could be more easily standardized. Consequently, integrating our process into the publishing workflow would represent a substantial advancement toward the broader goal of making scientific PDFs accessible to everyone.

\bibliographystyle{ACM-Reference-Format}
\bibliography{references}

\end{document}